\documentclass[conference]{IEEEtran}
\usepackage{array}
\usepackage{mathptmx}      
\usepackage{amsmath}       
\usepackage{amsthm}
\usepackage{pbox}
\usepackage{verbatim}
\usepackage{color}
\usepackage{amssymb}
\usepackage{enumerate}
\usepackage[utf8]{inputenc}
\usepackage[english]{babel}
\usepackage{cite}
\usepackage{todonotes}

\usepackage{txfonts}

\usepackage{graphicx}
\graphicspath{{./img/}}
\DeclareGraphicsExtensions{.pdf}
\usepackage{enumerate}

\newcommand{\myincludegraphics}[2]{
\begin{figure}
  \centering
  \includegraphics[scale=0.65]{img_#1}
  \caption{#2}
  \label{fig:#1}
\end{figure}
}

\newcommand{\myfigref}[1]{Figure~\ref{fig:#1}}
\newtheorem{definition}{Definition}
\newtheorem{theorem}{Theorem}

\begin{document}
\title{Data Attacks on Power System State Estimation: Limited Adversarial Knowledge vs Limited Attack Resources}

\author{\IEEEauthorblockN{Kaikai Pan\IEEEauthorrefmark{1},
Andr\'e Teixeira\IEEEauthorrefmark{2},
Milos Cvetkovic\IEEEauthorrefmark{1}
and Peter Palensky\IEEEauthorrefmark{1}} 
\IEEEauthorblockA{\IEEEauthorrefmark{1}Intelligent Electrical Power Grids\\
Faculty of EEMCS, Delft University of Technology,
Delft, The Netherlands}
\IEEEauthorblockA{\IEEEauthorrefmark{2}Engineering Systems and Services\\
Faculty of TPM, Delft University of Technology,
Delft, The Netherlands}}

\maketitle

\begin{abstract}

A class of data integrity attack, known as false data injection (FDI) attack, has been studied with a considerable amount of work. It has shown that with perfect knowledge of the system model and the capability to manipulate a certain number of measurements, the FDI attacks can coordinate measurements corruption to keep stealth against the bad data detection. However, a more realistic attack is essentially an attack with limited adversarial knowledge of the system model and limited attack resources due to various reasons. In this paper, we generalize the data attacks that they can be pure FDI attacks or combined with availability attacks (e.g., DoS attacks) and analyze the attacks with limited adversarial knowledge or limited attack resources. The attack impact is evaluated by the proposed metrics and the detection probability of attacks is calculated using the distribution property of data with or without attacks. The analysis is supported with results from a power system use case. The results show how important the knowledge is to the attacker and which measurements are more vulnerable to attacks with limited resources.

\end{abstract}

\IEEEpeerreviewmaketitle

\section{Introduction}

The integration of information and communication technology (ICT) and power systems makes the intelligent power grids typical cyber-physical systems. These systems are operated by means of complex distributed software systems which transmit information through wide and local area networks \cite{Teixeira2010}. Thus the intelligent power grids would be exposed to a large number of security threats \cite{ChenAbu-Nimeh2011, HongChenLiuEtAl2015}. The heterogeneity, diversity, and complexity of intelligent power grids may introduce new vulnerabilities that may lead to severe consequences \cite{Mo2012}. 

The State Estimation (SE) within modern energy management systems (EMS) is an instance of such dependency. It is supported by the Supervisory Control and Data Acquisition (SCADA) system for data delivery and provides important information to the EMS for power grids monitoring and control. SE uses measurements collected by the Remote Terminal Units (RTUs) in substations
and transmitted through the SCADA communication network to the control center. There is a built-in bad data detection (BDD) process in SE to detect erroneous measurements. The estimated state information is then processed by other applications in EMS such as optimal power flow (OPF) and Contingency Analysis (CA) to compute optimal control action while ensuring reliability and safety. The critical nature of SE highlights the importance of making it accurate and secure for power grid operations. In order to increase the security of SE and EMS, one needs to conduct vulnerability and attack impact assessment. Some of the literature has already tackled these problems. Vulnerability of SE to data integrity attack (e.g., false data injection (FDI) attack) is quantified by computing the attack resources needed by the adversary to keep stealth against the BDD \cite{Hug2012, Sandberg2010, TeixeiraSou2015}. The attack impacts of FDI attacks on SE, such as introduced estimate errors \cite{KosutJiaThomasEtAl2011}, potential economic loss in market operation \cite{XieMoSinopoli2011, Jia2014}, physical damaging like line overflows, are also well presented and analyzed. Besides, the FDI attacks with limited knowledge or limited resources are discussed by restricting the knowledge of the adversary to a part of the system model \cite{Rahman2012} or a subset of the network \cite{Zhang2016}, or restricting the capability of the attacker to manipulate the number of sensors \cite{KosutJiaThomasEtAl2011}. Our recent work \cite{Pan2016} extends the attack scenarios that the SE can be corrupted by FDI attacks and data availability attacks (e.g., Denial of Service (DoS) attacks) simultaneously.  

In this paper we aim to contribute in analyzing data attacks with limited adversarial knowledge and limited attack resources. Here the data attacks are ``generalized'' that can be pure integrity attacks (i.e. FDI attacks) or combined integrity and availability attacks. In order to achieve this, we introduce attack vectors for FDI attacks and availability attacks respectively and formulate attack strategies under both scenarios of limited knowledge and limited resources. To compare attacks under these two scenarios, we also propose attack impact metrics for evaluating attack impact on load estimate and provide methods to calculate detection probability of the attacks under these two scenarios. We show how important the knowledge of the system model is to the adversary and which measurements have the priority to be protected when attacks have limited resources. The analysis is supported with results from a case study.             

The outline of the paper is as follows. Section II details the state estimation techniques, the optimal attack strategy with full knowledge and attack resources. The attack impact metrics are also proposed to evaluate impact on load estimate. In Section III, the first scenario that attacks with limited adversarial knowledge is discussed. The method to calculate the detection probability of the attacks is discussed and a special case of limited knowledge scenario is specified. The second scenario that attacks with limited attack resources is presented in detail in Section IV, including the optimal data attacks with limited resources and the computation solution for solving it. Section V shows the results from the case study. The conclusion remarks are in Section VI.

\section{System Model and Data Attacks} 

In this section, we review the basic state estimation problem and discuss the optimal data attacks with perfect knowledge and full resources. The attack impact metrics are also formulated. 

\subsection{DC State Estimation and Bad Data Detection}

A power system model has a number of buses connected by transmission lines. The measurement data collected by RTUs includes line power flow measurements and bus power injection (generation, load) measurements. In the formulation, we use the DC power system model by neglecting the reactive power, line losses and assuming the voltage magnitudes to be constant. This model is commonly employed in security analysis of SE.

We assume that there are $n+1$ buses and $n_t$ transmission lines in the power network. Line power flow and bus power injection measurements are collected by RTUs in each substation. These $m$ power flow measurements are denoted by $z = [z_{1}, \ldots, z_{m}]^{T}$. The DC SE solves the following problem,
\begin{equation}
\label{eq: DC_SE}
z = \left[\begin{matrix} P_{1}WB^T\\ -P_{2}WB^T\\ P_{3}B_{0} W B^T\\ \end{matrix}\right]x + e := Hx + e,
\end{equation}
where $H \in \mathbb{R}^{m \times n}$ represents the system model, depending on the parameters of transmission lines (i.e., matrix $W$), the network topology (i.e., matrix $B_{0}$) and the placement of RTUs (i.e., matrices $P_{1}$, $P_{2}$, $P_{3}$). Here the topology is described by a directed incidence matrix $B_{0} \in \mathbb{R}^{(n+1) \times n_t}$ in which the directions of the lines can be arbitrarily specified \cite{TeixeiraSou2015}. Matrix $B \in \mathbb{R}^{n \times n_t}$ is the truncated incidence matrix with the row in $B_{0}$ corresponding to the reference bus removed. Matrix $W \in \mathbb{R}^{n_t \times n_t}$ is a diagonal matrix whose diagonal entries are the reciprocals of transmission line reactance. Matrices $P_1$, $P_2$ and $P_3$ are stacked by the rows of identity matrices, indicating whether a particular line power flow at the both sides of lines or bus power injection is measured. The total number of rows of $P_1$, $P_2$ and $P_3$ is $m$. The state vector $x = [x_{1},\ldots,x_{n}]^{T}$ refers to phase angles on each bus except the reference one and $e \sim \mathcal{N}(0 ,R)$ is the measurement noise vector of independent zero-mean Gaussian variables with covariance matrix $R = \mbox{diag}(\sigma_{1}^{2}, \ldots, \sigma_{m}^{2})$.

The state estimate $\mathit{\hat{x}}$ can by obtained using weighted least squares (WLS) estimate:
\begin{equation}
\label{eq: state_esti} 
\mathit{\hat{x}} = \mbox{arg} \min\limits_{x} (\mathit{z} - \mathit{Hx})^{T} \Sigma^{-1}(\mathit{z} - \mathit{Hx}),
\end{equation}
which can be solved as
\begin{equation}
\label{eq: DCSE_x} 
\hat{x} = (H^{T}R^{-1}H)^{-1}H^{T}R^{-1}z := Kz.
\end{equation}

To validate state estimates, the bad data detection schemes are used to detect erroneous measurements. The algorithms of the BDD within SE are based on the measurement residual
\begin{equation}
\label{rediual} 
r = z - H\hat{x} = (I-HK)z := Sz,
\end{equation}
where $r \in \mathbb{R}^{m}$ is the residual vector, $I \in \mathbb{R}^{m \times m}$ is an identity matrix, and $S$ is the so-called residual sensitivity matrix \cite{AburExposito2004}. The BDD is based on checking whether the $p$-norm of (weighted) measurement residual is below some threshold $\tau$. In this paper we choose the $J(\hat{x})$-test based BDD, which uses the quadratic function $J(\hat{x})=\lVert R^{-1/2}r \rVert_{2}^{2}$ to check if it follows the chi-squared distribution $\chi_{m-n}^{2}$. The BDD scheme becomes
\begin{equation}
\label{bdd_test}
\left\{
\begin{array}{rcl}
&\text{Good data, if}& \lVert R^{-1/2}r \rVert_{2}^{2} \leq \tau(\alpha), \\
&\text{Bad data, if}& \lVert R^{-1/2}r \rVert_{2}^{2} > \tau(\alpha), \\
\end{array}
\right.
\end{equation}
where $\tau(\alpha)$ is the threshold corresponding to the false alarm rate $\alpha$. Defining the probability distribution function (PDF) of $\chi_{m-n}^{2}$, $\tau(\alpha)$ can be obtained by solving    
\begin{equation}\label{dect_alpha}
\int_{0}^{\tau(\alpha)} {f(x)dx} = 1 - \alpha. 
\end{equation}

\subsection{Optimal Data Attacks with Perfect Knowledge and Full Resources}

An adversary aims to perturb the state estimate and keep stealth against the BDD, by tampering of RTUs, the SCADA system or even the SCADA master in the control center. To generalize the potential data attack, we assume that the attacker will use all tools available and can launch both data integrity and availability attacks. Corrupted by such data attacks, the measurement vector $z$ is changed into 
\begin{equation}\label{attacked_z}
\overline{z}:= (I - \mbox{diag}(d))z + a,
\end{equation}
where $a \in \mathbb{R}^{m}$ is the \emph{FDI attack vector}, $d \in \{0,1\}^{m}$ is the \emph{availability attack vector} and $I \in \mathbb{R}^{m \times m}$ is an identity matrix. Throughout this paper we define $diag(d)$ as the $m \times m$ diagonal matrix with the elements of vector $d$ on the main diagonal. To describe the number of attacked measurements needed by the adversary, we have the following definition of a $(k_{a}, k_{d})$-tuple attack, 
\begin{definition}[$(k_{a}, k_{d})$-tuple attack]
\label{def_Com}
A data attack with an FDI attack vector $a \in \mathbb{R}^{m}$ and an availability attack vector $d \in \{0,1\}^{m}$ described above is called a $(k_{a}, k_{d})$-tuple attack if 
$\| a \|_{0} = k_{a}$, $\| d \|_{0} = k_{d}$.  
\end{definition}
%

%
%

In our recent work \cite{Pan2016}, we showed that if the attacker has the perfect knowledge of the system model (i.e., $H$) and can manipulate certain number of measurements with full attack resources to perform a $(k_{a}, k_{d})$-tuple attack, 
it can keep stealth against the BDD by introducing the two attack vectors $a$ and $d$ satisfying $a=(I - \mbox{diag}(d))Hc$ where $c \in \mathbb{R}^{n}$ is non-zero. Under a such $(k_{a}, k_{d})$-tuple attack, we define the matrix of the system model and the noise vector as a function of the attack vector $d$,
\begin{equation}
H_{d}:= (I - \mbox{diag}(d))H, \quad e_{d}:= (I - \mbox{diag}(d))e.
\end{equation}
where $H_{d}$ denotes the model of the remaining measurements and it is obtained from $\mathit{H}$ by replacing some rows with zero row vectors due to availability attacks on these measurements, $e_{d}$ is the noise vector for remaining measurements and the entries of it are zero if the corresponding measurements are unavailable. According to the formulation of matrix $K$ in \eqref{eq: DCSE_x} and matrix $S$ in \eqref{rediual}, we can also have
\begin{equation}
K_{d}:= (H_{d}^{T}R^{-1}H_{d})^{-1}H_{d}^{T}R^{-1},
\end{equation}
\begin{equation}
\label{eq: T0} 
S_{d} := I  - H_{d}K_{d}. 
\end{equation}

An optimal attack pursue minimum attack resources. To simplify the discussion, we assume that an optimal attack can have the minimum attack resources when it needs to corrupt the minimum number of measurements. For the adversary with the capability to manipulate a certain number of measurements and perfect knowledge of the system model, we formulate the optimal attack strategy as the following optimization problem,
\begin{subequations}\label{attackresources_com_o}
\begin{align}
(c^{*},d^{*}):= \mbox{arg} \min\limits_{c,d}\quad& \lVert a \rVert_{0} + \lVert d \rVert_{0} \nonumber\\
\mbox{s.t.}\quad& a = H_{d}c, \label{eq:11a}\\
 & H_{d} =(I - \mbox{diag}(d))H, \label{eq:11b}\\ 
 & a(j) = \mu, \label{eq:11c}\\
 & d(i) \in \{0,1\}\quad\mbox{for all } i. \nonumber
\end{align}
\end{subequations}

Here in \eqref{eq:11b} we assume $a(j) = \mu$ where $\mu$ is the non-zero attack magnitude on the target measurement $j$. For measurement $j$, the optimal attack with attack vectors $d^{*}$ and $a^{*}=H_{d^{*}}c^{*}$ has the minimum number of measurements to corrupt and correspondingly has the minimum attack resources. To solve the optimization problem above, we propose a computation solution which uses the big M method: 

\begin{subequations}\label{attackresources_com_bigM} 
\begin{align}
(c^{*}, w^{*}, d^{*}):= \mbox{arg} \min\limits_{c,w,d}\quad &\sum\limits_{i=1}^{m} w(i) + \sum\limits_{k=1}^{m} d(k) \nonumber\\
\mbox{s.t.}\quad& Hc \leq M(w+d), \label{eq:12a}\\
 & -Hc \leq M(w+d),  \\
 & H(j,:)c = \mu,  \label{eq:12c}\\
 & w(i) \in \{0,1\}\quad\mbox{for all } i, \label{eq:12d}\\
 & d(k) \in \{0,1\}\quad\mbox{for all } k, \label{eq:12e}
\end{align}
\end{subequations}
where $w, d \in \{0,1\}^{m}$ in \eqref{eq:12d} and \eqref{eq:12e}. $w(i) = 1$ and $d(k) = 1$ means that FDI attack and data availability attack take place on measurement $i$ and $k$ respectively. The following theorem, which is adopted from our work \cite{Pan2016}, states that the optimal solution to \eqref{attackresources_com_o} can be exactly obtained by solving \eqref{attackresources_com_bigM}.
\begin{theorem}
For any measurement index $j\in\{1,\dots,m\}$ and non-zero $\mu$, let ($c^{*}$, $w^{*}$, $d^{*}$) be an optimal solution to \eqref{attackresources_com_bigM}. Then an optimal solution to \eqref{attackresources_com_o} can be computed as ($c^{*}$, $d^{*}$). 
\end{theorem}
%

\subsection{Attack Impact Metric}

As the work in \cite{Liang2016}\cite{Teixeira2012} shows, the OPF application uses the load estimate as the inputs provided by SE. If data attacks take place and pass the BDD, the load estimate get perturbed and it will influence the control actions in the next time interval. Therefore, we consider the impact metric as a function of the bias introduced by the attack on the load estimate.

Assuming that there are $m_{inj}$ injection (with load) measurements, we consider the impact on the errors of power injection (with load) estimate, which can be described as
%
\begin{equation}\label{err_inj}
\epsilon = \hat{z}_{inj,a,d} - z_{inj},
\end{equation}
where $z_{inj} \in \mathbb{R}^{m_{inj}} $ is the vector of power injection (with load) measurements without attacks and $\hat{z}_{inj, a, d} \in \mathbb{R}^{m_{inj}}$ is the vector of estimated injection (with load) measurements under a $(k_{a}, k_{d})$-tuple attack. We can further obtain
\begin{equation}
\epsilon = H_{inj} \hat{x}_{a,d} - (H_{inj}x+e_{inj}),
\end{equation}
where $\hat{x}_{a,d}=K_{d}\overline{z}=x+K_{d}e_{d}+K_{d}a$, $H_{inj} \in \mathbb{R}^{m_{inj} \times n}$ denotes the submatrix of $H$ by keeping the rows corresponding to injection (with load) measurements, i.e. $H_{inj} = M_{inj}H$ where $M_{inj} \in \mathbb{R}^{m_{inj} \times m}$ is the incidence matrix for selecting the rows in $H$ corresponding to injection (with load) measurements, $e_{inj} \in \mathbb{R}^{m_{inj}}$ is the noise vector for injection (with load) measurements. Thus $\epsilon = H_{inj} K_{d}a - M_{inj}S_{d}e$. The expected value of injection (with load) estimate errors is
\begin{equation}\label{exp_err}
\mathbb{E}(\epsilon) = H_{inj} K_{d}a.
\end{equation}
%
%
%
We have the following definition for the attack impact metric,
\begin{definition}\label{def_attackimpact}
The impact metric $\mathbf{I}(a,d)$ for quantifying attack impact of a $(k_{a}, k_{d})$-tuple attack with attack vectors $a$ and $d$ on load estimate is defined as the 2-norm of $H_{inj} K_{d}a$, i.e. $\mathbf{I}(a,d) = \lVert H_{inj} K_{d}a \rVert_{2}$.
\end{definition}

\section{Scenario 1: Attacks with Limited Adversarial Knowledge}

In this section, we consider the first scenario that the adversary has limited knowledge of the system model and discuss how this would affect the detectability of data attacks.

\subsection{Perturbed Model Known by the Attacker}

In Section II-B, the attacks are assumed to have perfect knowledge of the system model. This requires knowledge on topology matrix $B_{0}$, the line parameter matrix $W$ and the RTU placement matrix $P_{1}$, $P_{2}$, $P_{3}$. Such knowledge can be obtained by recording and analyzing data sent from RTUs using statistical methods. However, due to restricted access to the power grid, errors in data collection and analyzing may essentially result in an attack with limited knowledge of the system model. We denote the perturbed system model known by the attacker as $\tilde{H}$, such that 
\begin{equation}
\label{err_H}
\tilde{H} = H + \Delta H, 
\end{equation}
where $\Delta H \in \mathbb{R}^{m \times n}$ denotes the part of model uncertainty because of the issues discussed above.

\subsection{Detection Probability of Attacks}

When the measurements are corrupted by a $(k_{a}, k_{d})$-tuple attack, the measurement residual $r(a,d)$ can be written as
\begin{equation}\label{residual_attack}
r(a,d) = S_d\overline{z}=S_{d}e_{d} + S_{d}a.
\end{equation}
%

As discussed in Section II-B, when the vectors of the $(k_{a}, k_{d})$-tuple attack satisfy $a=H_{d}c$, the residual $r(a,d) = S_{d}e_{d} + S_{d}H_{d}c = S_{d}e_{d}$ due to $S_{d}H_{d}=0$. Then the residual is not affected by $a$ and no increase of alarms would triggered in the BDD since the BDD would treat the measurements attacked by availability attacks as a case of data missing. However, when the attacker has limited knowledge of the system model, the attack vector $a$ becomes $a=(I - \mbox{diag}(d))\tilde{H}c:=\tilde{H}_{d}c$ and $S_{d}a$ may be non-zero in this case. The measurement residual is incremented and the attack can be detected with some probability. In the following we show how the detection probability can be calculated.
 
Note that the quadratic cost function with a $(k_{a}, k_{d})$-tuple attack becomes $J_{a,d}(\hat{x})=\lVert R^{-1/2}r(a,d) \rVert_{2}^{2}$. We can further obtain $J_{a,d}(\hat{x}) = \lVert R^{-1/2}S_{d}e_{d} + R^{-1/2}S_{d}a \rVert_{2}^{2}$. Here the mean of $( R^{-1/2}S_{d}e_{d} + R^{-1/2}S_{d}a)$ becomes non-zero $R^{-1/2}S_{d}a$ incremented by the attack. 
Under the $(k_{a}, k_{d})$-tuple attack, $J_{a,d}(\hat{x})$ has a \emph{generalized non-central chi-squared distribution} with $m - n - k_{d}$ degrees of freedom. We use $J_{a,d}(\hat{x})$ as an approximation of having the \emph{non-central chi-squared distribution} $\chi_{m-n-k_{d}}^{2}(\lVert R^{-1/2}S_{d}a \rVert_{2}^{2})$ to calculate the detection probability, where $\lambda_{a,d} = \lVert R^{-1/2}S_{d}a \rVert_{2}^{2}$ is the non-centrality parameter. In \cite{Pan2017}, we have validated such approximation using empirical results from Monte Carlo simulation. It implies that data attacks with limited adversarial knowledge would increase the possibility to trigger alarms in BDD due to the introduced non-zero non-centrality parameter. We can get
\begin{equation}\label{nchi_pdf}
\int_{0}^{\tau_{d}(\alpha)} {f_{\lambda_{a,d}}(x)dx} = 1 - \delta_{a,d}, 
\end{equation}
where $f_{\lambda_{a,d}}(x)$ is the PDF of $\chi_{m-n-k-d}^{2}(\lVert R^{-1/2}S_{d}a \rVert_{2}^{2})$, $\tau_{d}(\alpha)$ is the threshold set in the BDD using \eqref{dect_alpha} but with the PDF of $\chi_{m-n-k_{d}}^{2}$, 
and $\delta_{a,d}$ is the detection probability of the $(k_{a}, k_{d})$-tuple attack. In order to keep stealth against the BDD, the attacker has to minimize $\delta_{a,d}$ as close to the false alarm rate $\alpha$ as possible.

\subsection{A Special Case of Limited Adversarial Knowledge}

The model uncertainty defined in \eqref{err_H} is ``general'' and can be any values. An interesting analysis is to understand what the model uncertainty $\Delta H$ could be to the adversary. In particular, we are interested in the case where the adversary knows the exact topology of the power network and the placement of RTUs, but has limited information of the line parameters. This can be expected due to various practical reasons as explained in \cite{Rahman2012}, e.g., limited access to the knowledge of exact position of the tap changer and the exact length of the transmission line and type of the conductor being used. Thus the perturbed system model known by the adversary becomes
\begin{equation}
\label{structured_uncertainty}
\tilde{H} = P \left[\begin{matrix} (W+\Delta W)B^T\\ -(W+\Delta W)B^T\\ B_{0} (W+\Delta W) B^T\\ \end{matrix}\right],
\end{equation}
%
where $\Delta W \in \mathbb{R}^{n_{t} \times n_{t}}$ denotes the error on estimate of transmission line reactance.      

\section{Scenario 2: Attacks with Limited Attack Resources}

In this section we consider the second scenario where the adversary has limited attack resources but still targets to keep stealth against the BDD and have maximum attack impact.

\subsection{Optimal Data Attacks with Limited Attack Resources}

The attack policies in Section II-B for the $(k_{a}, k_{d})$-tuple attacks follow the linear model and the adversary is assumed to be able to attack a certain number of measurements, i.e., $k_a + k_d \geq \min \| a^{*} \|_{0} + \| d^{*} \|_{0}$ where attack vectors $d^{*}$ and $a^{*}=H_{d^{*}}c^{*}$ are obtained by solving \eqref{attackresources_com_bigM} for any measurement index $j$. 
Now we consider the following scenario that the attacker has limited attack resources that $k_a + k_d < \min \| a^{*} \|_{0} + \| d^{*} \|_{0}$ and thus can not follow the linear attack policies above. For the sake of comparison, in this scenario we assume that the attacker has full knowledge of the system model and enough computational capability. In the following we construct the optimal attack strategy for the $(k_{a}, k_{d})$-tuple attacks with limited attack resources, which is also the worst case from the perspective of system operation. 

The objective of the attack is to gain an ability to introduce error on load estimate. The adversary tries to achieve the goal by maximizing the impact metrics, i.e., maximizing $\lVert H_{inj}K_{d}a \rVert_{2}$ for a $(k_a, k_d)$-tuple attack.

An optimal attack also targets to keep stealth against the BDD, or at least minimize the detection probability. In the case that the adversary has limited attack resources, the attacks can be detected since the term $S_{d}a$ in \eqref{residual_attack} may be non-zero. We assume there exists an upper limit of detection probability $\overline{\delta}$ which is acceptable for the adversary. Thus according to Subsection III-B, for the $(k_{a}, k_{d})$-tuple attack, we can obtain 
\begin{equation}\label{nchi_pdf_pp}
\int_{0}^{\tau_{d}(\alpha)} {f_{\lambda_{a,d}}(x)dx} \geq 1 - \overline{\delta}, 
\end{equation}
where the non-centrality parameter is $\lambda_{a,d} = \lVert R^{-1/2}S_{d}a \rVert_{2}^{2}$ and $f_{\lambda_{a,d}}(x)$ is the PDF of $\chi_{m-n-k_{d}}^{2}(\lVert R^{-1/2}S_{d}a \rVert_{2}^{2})$. For a given $\overline{\delta}$, the non-centrality parameter satisfies $\lambda_{a,d} = \lVert R^{-1/2}S_{d}a \rVert_{2}^{2} \leq \overline{\varepsilon}_{a,d}$ where $\overline{\varepsilon}_{a,d}$ can be determined using \eqref{nchi_pdf_pp}.

Based on the aforementioned intuition, we consider the following optimization problem for the optimal attack strategy of the $(k_{a}, k_{d})$-tuple attack under the relaxation on the assumption of attack resources,
\begin{align}\label{attackimpact_limitedR}
\begin{split}
\max\limits_{a,d} \quad& \lVert H_{inj}K_{d}a \rVert_{2}^{2} \\
\mbox{s.t.}\quad& \lVert R^{-1/2}S_{d}a \rVert_{2}^{2} \leq \overline{\varepsilon}_{a},\\
 & \lVert a \rVert_{0} + \lVert d \rVert_{0} \leq \overline{R}.
\end{split}
\end{align}
where $\overline{R}$ denotes the maximum number of measurements that the attacker can manipulate.
 
\subsection{Computation Solution}

The optimization problem of \eqref{attackimpact_limitedR} for a $(k_a, k_d)$-tuple attack is non-convex and difficult to solve without the specifications of the attack vectors \cite{KosutJiaThomasEtAl2011}. We consider to add more constraints on the attack vectors by setting the measurements which could be attacked to be determined for a given $k_a$ and $k_d$ which satisfy $k_{a} + k_{d} \leq \overline{R}$. Thus for this specifications of the non-zero entries  in attack vectors $a$ and $d$, the above problem can be equivalent to solve the following one,
\begin{align}\label{attackimpact_limitedR2}
\begin{split}
\min\limits_{a} \quad& \overline{\varepsilon}_{a,d} \varphi \\
\mbox{s.t.}\quad& Q -  \varphi W \leq 0,\\
\end{split}
\end{align}
where $Q = Qs^TQs$, $W=Ws^TWs$, and for a givn $d$, $Qs$, $Ws$ are the submatrices of $H_{inj}K_{d}$ and $R^{-1/2}S_{d}$ corresponding to the non-zero entries of attack vector $a$. \eqref{attackimpact_limitedR2} also implies that $\varphi$ is the maximum generalized eigenvalue of the matrix pair $(Q, W)$, i.e., $\varphi=\lambda_{max}(Q,W)$ \cite{Boyd1993}.

It should be noted that though the attack vectors are constrained in order to solve the optimization problem, the optimal attacks for any given $k_{a}$ and $k_{d}$ which satisfy $k_{a} + k_{d} \leq \overline{R}$ can still be obtained using exhaustive search over all possible attacked measurement sets. Some computationally efficient algorithms can also be developed to solve \eqref{attackimpact_limitedR}. We leave this for future work. 

\section{Case Study}

In this section we apply the attack scenarios of limited adversarial knowledge and limited attack resources to the IEEE 14-bus system use case. Full measurements placement is employed that power flow measurements are placed on all the buses and transmission lines to provide large redundancy. The per-unit system is used and the power base is $100 MW$. The power flow measurements are generated by the DC model with Gaussian noise ($\sigma_i = 0.02$ for all measurements). For the limited knowledge scenario, we assume that 
the adversary knows the exact topology of the system but has an estimated line parameters with errors up to $\pm$10$\%$, $\pm$20$\%$ and $\pm$30$\%$.

With different levels of error on estimation of line parameters, the detection probability of the attacks can be obtained according to \eqref{nchi_pdf}. We pick such case that measurement $j=9$. According to the optimal solution of \eqref{attackresources_com_bigM} w.r.t. $\tilde{H}$ in \eqref{structured_uncertainty}, a number of 11 measurements need to be manipulated by the attacker. \myfigref{Figure1_detectionprob} shows the detection probability of the (5,6)-tuple attacks targeting on these 11 measurements. The x-axis indicates the attack magnitude $\mu$ on measurement $j$ using the attack vector $a = \tilde{H}_{d}c$. We can see that the detectability of attack is intimately related to the error on estimation of line parameters. With larger model uncertainty in building attack vectors, the  attack has higher possibility to be detected by the BDD. The detection probability becomes much higher when the error on estimation of line parameters increases. This implies that in order to keep stealth, the adversary do need good knowledge of the system model.  

\myincludegraphics{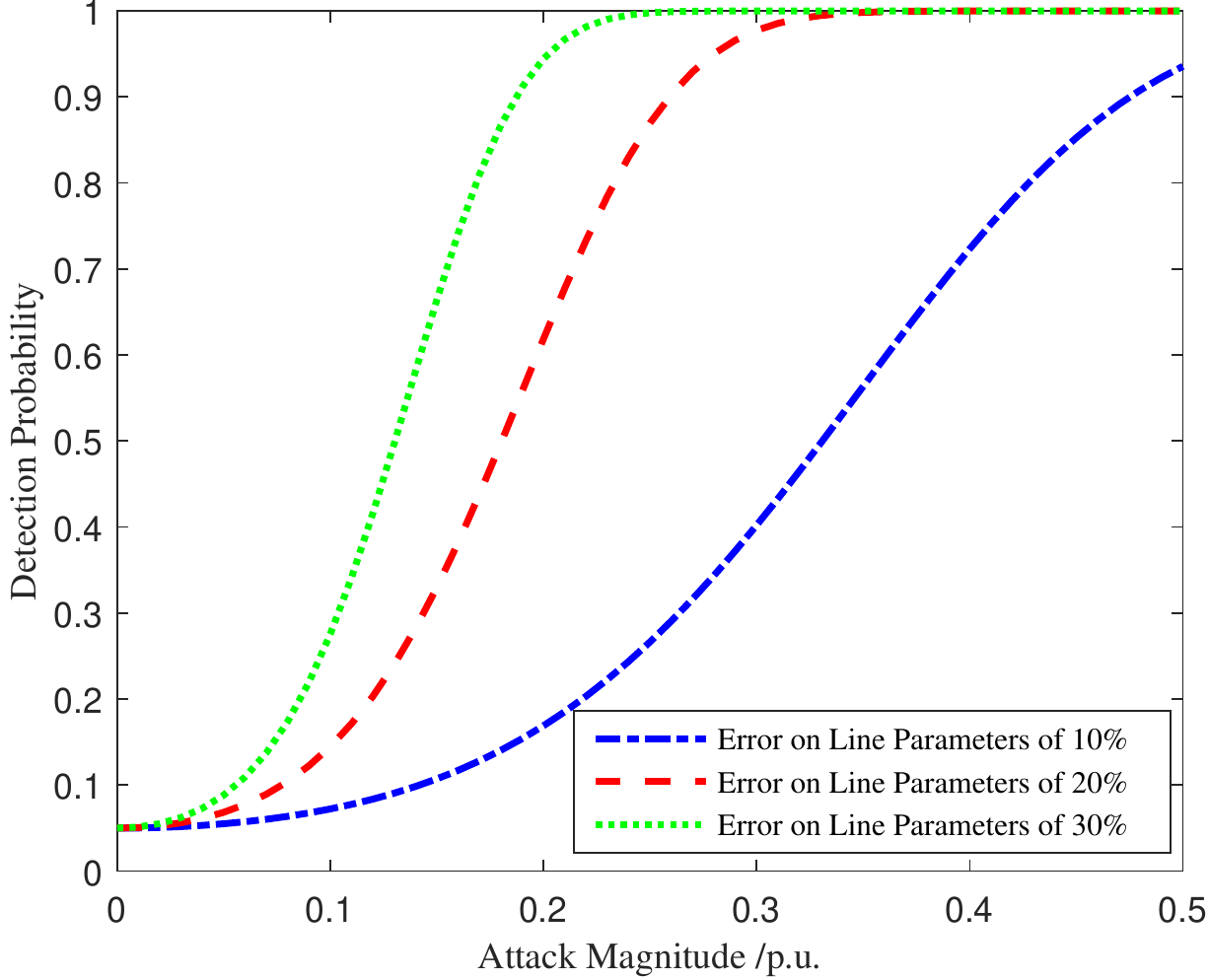}{ The detection probability is plotted versus the attack magnitude $\mu$ on measurement $j = 9$. The attacks are under there levels of error on estimation of line parameters: $\pm$10$\%$, $\pm$20$\%$ and $\pm$30$\%$. The false alarm rate is set to be $\alpha = 0.05$.} 

Next the simulations are conducted in the scenario where the adversary has limited attack resources but full knowledge of the system and the optimal attack strategy described in \eqref{attackimpact_limitedR} is used. The measurements that could be manipulated are specified on a determined measurement set which is the same as the one in the previous case, i.e., the set with 11 measurements containing measurement $j=9$. If all the measurements in this set are corrupted with enough attack resources, the attack with full knowledge can perform the optimal attack strategy in \eqref{attackresources_com_o} and keep stealth. However, in this scenario the attacker could only corrupt part of the measurements in the set thus can be observed by the BDD. Using \eqref{attackimpact_limitedR}, we compare the attack scenarios where the attacker has limited resources to corrupt part of the measurements but full knowledge and has limited knowledge but full resources to corrupt all the measurements in this measurement set, as shown in \myfigref{Figure2_limitedKnowledgevsResources}. We can see that though the attacker with limited resources can manipulate fewer measurements for (10,0)-tuple attack and (4,6)-tuple attack with full knowledge, they can have larger impact metrics comparing with the (11,0)-tuple attack and (5,6)-tuple attack with limited knowledge of the system. This implies the importance of the knowledge of system model. From the system operator's view, the system model kept in the database of SCADA should be protected well. Besides this also can be used to implement mitigation schemes by misleading attacks to use perturbed or even faked system model thus making them detectable. 

\myincludegraphics{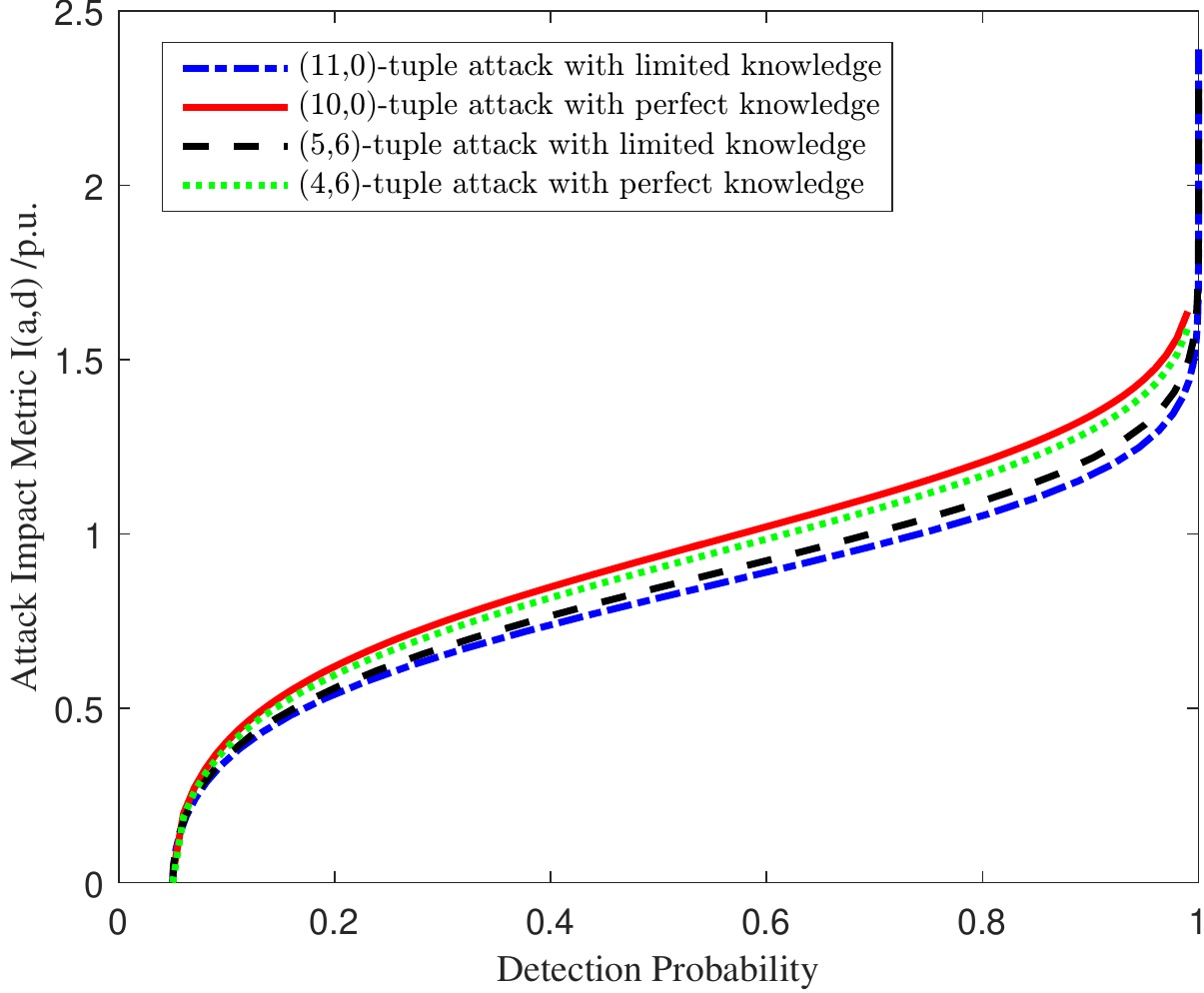}{The attack impact metric is plotted versus the detection probability under the attacks with limited knowledge or resources respectively. For the limited knowledge scenario, error on line parameters of $\pm$20$\%$ is employed. The false alarm rate is set to be $\alpha = 0.05$.} 

Then the scenario that the attacker can corrupt even fewer measurements is considered. Such $(k_{a}, k_{d})$-tuple attack becomes a ``sparse'' attack. All of the possible attacked measurement sets can be examined and the worst case with largest attack impact metrics is selected. Table~\ref{tab1} gives the results of the ``sparse'' optimal attacks. With a given $\overline{\delta}$ and $k_a$, $k_d$, the optimal attacks with largest attack impact metrics are obtained. Here the measurements index denotes the measurement set manipulated by the attacker. Index 7 and 27 are line power flow measurements on branch 7 (from bus 4 to bus 5) and 42, 44, 45 are bus power injection measurements on bus 2, 4, 5. We can see that with more measurements corrupted by FDI attack, the maximum impact metrics on errors of load estimate can be larger. The results also indicate that the measurement set (e.g., line power flow measurements on branch 7 and injection measurements on bus 2,4,5) are vulnerable to the ``sparse'' attacks from the view of the system operation and has the priority of be equipped with mitigation schemes.

\begin{table}[t]
  \centering
   \caption{Sparse Optimal Attacks with Limited Resources}\label{tab1}
    \begin{tabular}{|p{0.12\textwidth}|p{0.06\textwidth}|p{0.07\textwidth}|p{0.07\textwidth}|}
    \hline
    Attacks & Index & $\mathbf{I}(a,d)$ /p.u. ($\overline{\delta}=0.1$) & $\mathbf{I}(a,d)$ /p.u. ($\overline{\delta}=0.2$)   
    \\
    
    \hline
    (3,0)-tuple attack & (7,27,45) &0.0585 & 0.0898 
    \\

    \hline{}
    (2,0)-tuple attack & (42,45) & 0.0440 & 0.0676 
    \\

    \hline{}
    (1,0)-tuple attack & (44) & 0.0354 & 0.0544 
    \\

    \hline
   (2,1)-tuple attack & (7,27,45) & 0.0574 & 0.0881 
    \\
    
    \hline
   (1,2)-tuple attack & (7,27,45) & 0.0551 & 0.0847 
    \\
 
    \hline
   (1,1)-tuple attack & (7,44) & 0.0412 & 0.0633 
    \\
    
    \hline
  \end{tabular}
\end{table}

\section{Conclusion}

In this paper, we consider more realistic attacks with limited adversarial knowledge and limited attack resources. The attack is also generalized to include both data integrity and data availability attacks. We show that the detection probability of attacks increase when the error on parameter estimation increase for the attack. The optimal attacks with limited attack resources but full knowledge can be more damaging than the ones with limited knowledge but enough attack resources according to the detection probability and attack impact metrics, which implies the importance of the knowledge of the system model to the attack. The results also suggest which measurements are more vulnerable to ``sparse'' data attacks that need to be protected with priority. Future work includes the computationally efficient algorithms, using
various attack cost on different measurements, mitigation schemes, etc.  


\bibliographystyle{IEEEtran}
\bibliography{IEEEabrv,Literature}

\begin{thebibliography}{10}
\providecommand{\url}[1]{#1}
\csname url@samestyle\endcsname
\providecommand{\newblock}{\relax}
\providecommand{\bibinfo}[2]{#2}
\providecommand{\BIBentrySTDinterwordspacing}{\spaceskip=0pt\relax}
\providecommand{\BIBentryALTinterwordstretchfactor}{4}
\providecommand{\BIBentryALTinterwordspacing}{\spaceskip=\fontdimen2\font plus
\BIBentryALTinterwordstretchfactor\fontdimen3\font minus
  \fontdimen4\font\relax}
\providecommand{\BIBforeignlanguage}[2]{{%
\expandafter\ifx\csname l@#1\endcsname\relax
\typeout{** WARNING: IEEEtran.bst: No hyphenation pattern has been}%
\typeout{** loaded for the language `#1'. Using the pattern for}%
\typeout{** the default language instead.}%
\else
\language=\csname l@#1\endcsname
\fi
#2}}
\providecommand{\BIBdecl}{\relax}
\BIBdecl

\bibitem{Teixeira2010}
A.~Teixeira, S.~Amin, H.~Sandberg, K.~H. Johansson, and S.~S. Sastry, ``Cyber
  security analysis of state estimators in electric power systems,'' in
  \emph{Proc. 49th IEEE Conf. Decision and Control (CDC)}, Dec. 2010, pp.
  5991--5998.

\bibitem{ChenAbu-Nimeh2011}
T.~M. Chen and S.~Abu-Nimeh, ``Lessons from stuxnet,'' \emph{Computer},
  vol.~44, no.~4, pp. 91--93, 2011.

\bibitem{HongChenLiuEtAl2015}
J.~Hong, Y.~Chen, C.-C. Liu, and M.~Govindarasu, ``Cyber-physical security
  testbed for substations in a power grid,'' in \emph{Cyber Physical Systems
  Approach to Smart Electric Power Grid}.\hskip 1em plus 0.5em minus
  0.4em\relax Springer, 2015, pp. 261--301.

\bibitem{Mo2012}
Y.~Mo, T.~H.-J. Kim, K.~Brancik, D.~Dickinson, H.~Lee, A.~Perrig, and
  B.~Sinopoli, ``Cyber--physical security of a smart grid infrastructure,''
  \emph{Proceedings of the IEEE}, vol. 100, no.~1, pp. 195--209, 2012.

\bibitem{Hug2012}
G.~Hug and J.~A. Giampapa, ``Vulnerability assessment of {AC} state estimation
  with respect to false data injection cyber-attacks,'' \emph{IEEE Transactions
  on Smart Grid}, vol.~3, no.~3, pp. 1362--1370, Sep. 2012.

\bibitem{Sandberg2010}
H.~Sandberg, A.~Teixeira, and K.~H. Johansson, ``On security indices for state
  estimators in power networks,'' in \emph{First Workshop on Secure Control
  Systems (SCS), Stockholm}, 2010.

\bibitem{TeixeiraSou2015}
A.~Teixeira, K.~C. Sou, H.~Sandberg, and K.~H. Johansson, ``Secure control
  systems: A quantitative risk management approach,'' \emph{IEEE Control
  Systems}, vol.~35, no.~1, pp. 24--45, 2015.

\bibitem{KosutJiaThomasEtAl2011}
O.~Kosut, L.~Jia, R.~J. Thomas, and L.~Tong, ``Malicious data attacks on the
  smart grid,'' \emph{IEEE Transactions on Smart Grid}, vol.~2, no.~4, pp.
  645--658, 2011.

\bibitem{XieMoSinopoli2011}
L.~Xie, Y.~Mo, and B.~Sinopoli, ``Integrity data attacks in power market
  operations,'' \emph{IEEE Transactions on Smart Grid}, vol.~2, no.~4, pp.
  659--666, 2011.

\bibitem{Jia2014}
L.~Jia, J.~Kim, R.~J. Thomas, and L.~Tong, ``Impact of data quality on
  real-time locational marginal price,'' \emph{IEEE Transactions on Power
  Systems}, vol.~29, no.~2, pp. 627--636, Mar. 2014.

\bibitem{Rahman2012}
M.~A. Rahman and H.~Mohsenian-Rad, ``False data injection attacks with
  incomplete information against smart power grids,'' in \emph{Global
  Communications Conference (GLOBECOM), 2012 IEEE}.\hskip 1em plus 0.5em minus
  0.4em\relax IEEE, 2012, pp. 3153--3158.

\bibitem{Zhang2016}
J.~Zhang, Z.~Chu, L.~Sankar, and O.~Kosut, ``False data injection attacks on
  power system state estimation with limited information,'' in \emph{Power and
  Energy Society General Meeting (PESGM), 2016}.\hskip 1em plus 0.5em minus
  0.4em\relax IEEE, 2016, pp. 1--5.

\bibitem{Pan2016}
K.~Pan, A.~M.~H. Teixeira, M.~Cvetkovic, and P.~Palensky, ``Combined data
  integrity and availability attacks on state estimation in cyber-physical
  power grids,'' in \emph{Proc. IEEE Int. Conf. Smart Grid Communications
  (SmartGridComm)}, Nov. 2016, pp. 271--277.

\bibitem{AburExposito2004}
A.~Abur and A.~G. Exposito, \emph{Power system state estimation: theory and
  implementation}.\hskip 1em plus 0.5em minus 0.4em\relax CRC press, 2004.

\bibitem{Liang2016}
J.~Liang, L.~Sankar, and O.~Kosut, ``Vulnerability analysis and consequences of
  false data injection attack on power system state estimation,'' \emph{IEEE
  Transactions on Power Systems}, vol.~31, no.~5, pp. 3864--3872, Sep. 2016.

\bibitem{Teixeira2012}
A.~Teixeira, H.~Sandberg, G.~Dan, and K.~H. Johansson, ``Optimal power flow:
  Closing the loop over corrupted data,'' in \emph{Proc. American Control Conf.
  (ACC)}, Jun. 2012, pp. 3534--3540.

\bibitem{Pan2017}
K.~Pan, A.~Teixeira, M.~Cvetkovic, and P.~Palensky, ``Cyber risk analysis of
  combined data attacks against power system state estimation,'' unpublished.

\bibitem{Boyd1993}
S.~Boyd and L.~El~Ghaoui, ``Method of centers for minimizing generalized
  eigenvalues,'' \emph{Linear algebra and its applications}, vol. 188, pp.
  63--111, 1993.

\end{thebibliography}
%

\end{document}